\documentclass[sigconf]{acmart}

\AtBeginDocument{%
  \providecommand\BibTeX{{%
    \normalfont B\kern-0.5em{\scshape i\kern-0.25em b}\kern-0.8em\TeX}}}

\setcopyright{acmcopyright}
\copyrightyear{2018}
\acmYear{2018}
\acmDOI{XXXXXXX.XXXXXXX}

\acmConference[Conference acronym 'XX]{Make sure to enter the correct
  conference title from your rights confirmation emai}{June 03--05,
  2018}{Woodstock, NY}
\acmPrice{15.00}
\acmISBN{978-1-4503-XXXX-X/18/06}
\usepackage{subfigure,algorithm,algorithmic,multirow}

\begin{document}

\title[Log-Spectral Matching GAN]{Log-Spectral Matching GAN: PPG-based Atrial Fibrillation Detection can be Enhanced by GAN-based Data Augmentation with Integration of Spectral Loss}

\author{Cheng Ding}
\affiliation{%
  \institution{Department of Biomedical Engineering, Georgia Institute of Technology \& Emory University}
  \city{Atlanta, GA}
  \country{USA}}

\author{Ran Xiao}
\affiliation{%
  \institution{School of Nursing, Duke University}
  \city{Durham, NC}
  \country{USA}
}

\author{Duc Do}
\affiliation{%
 \institution{David Geffen School of Medicine, University of California at Los Angeles}
 \city{Los Angeles, CA}
 \country{USA}}

\author{David Scott Lee}
\affiliation{%
  \institution{Department of Otolaryngology, Washington University in St. Louis}
  \city{St Louis, MO}
  \country{USA}}

\author{Shadi Kalantarian}
\affiliation{%
  \institution{School of Medicine, University of California San Francisco}
  \city{San Francisco, CA}
  \country{USA}}

\author{Randall J Lee}
\affiliation{%
  \institution{School of Medicine, University of California San Francisco}
  \city{San Francisco, CA}
  \country{USA}}
  
\author{Xiao Hu}
\affiliation{%
  \institution{School of Nursing, Emory University}
  \institution{Department of Biomedical Informatics, School of Medicine, Emory University}
  \city{Atlanta, GA}
  \country{USA}}

\renewcommand{\shortauthors}{Cheng Ding and Ran Xiao, et al.}

\begin{abstract}
Photoplethysmography (PPG) is a ubiquitous physiological measurement that detects beat-to-beat pulsatile blood volume changes and hence has a potential for monitoring cardiovascular conditions, particularly in ambulatory settings. A PPG dataset that is created for a particular use case is often imbalanced, due to a low prevalence of the pathological condition it targets to predict and the paroxysmal nature of the condition as well. To tackle this problem, we propose log-spectral matching GAN (LSM-GAN), a generative model that can be used as a data augmentation technique to alleviate the class imbalance in a PPG dataset to train a classifier. LSM-GAN utilizes a novel generator that generates a synthetic signal without a up-sampling process of input white noises, as well as adds the mismatch between real and synthetic signals in frequency domain to the conventional adversarial loss. In this study, experiments are designed focusing on examining how the influence of LSM-GAN as a data augmentation technique on one specific classification task - atrial fibrillation (AF) detection using PPG. We show that by taking spectral information into consideration, LSM-GAN as a data augmentation solution can generate more realistic PPG signals. The code of LSM-GAN is available at \url{https://github.com/chengding0713/Log-Spectral-matching-GAN}.
\end{abstract}

\maketitle

\section{Introduction}
\label{sec1}
It is reported that approximately 1.5\% - 2\% of the general adult population are affected by atrial fibrillation (AF) \cite{1}, and the prevalence of AF is expected to increase over the years due to an aging population. If left untreated, AF confers various significant health risks. AF is linked to a 5-fold increase in the risk of ischemic stroke, a 3-fold increase in the risk of heart failure, and a 2-fold increase in the risk of mortality from heart disease \cite{1}. Therefore, it marks great clinical and economic significance to have an affordable, portable, and continuous AF screening tool that patients with AF can access at scale. In the past, AF detection has been mainly relying on the analysis and interpretation of electrocardiogram (ECG). Recent advancement in wearable technologies, such as fitness bands and smartwatches, offers convenient and continuous recordings of photoplethysmography (PPG), which demonstrates to be a potential alternative to ECG for AF detection \cite{2,3}. Current wearables with continuous monitoring of PPG offer many benefits, such as friendly user interface, low cost, and portability, making it a promising platform to achieve an AF screening tool accessible to the general population. Therefore, PPG has garnered tremendous research interest in recent years for a reliable and accurate AF detection solution.  For PPG-based AF detection, it has been shown that deep learning (DL) algorithms achieve better results than traditional machine learning algorithms \cite{4,5,6,7}. Recent studies have shown the benefits using a balanced sample setup to train AF classifiers \cite{5,6,7,8,9,10}. However, it remains a challenge to obtain a balanced dataset with a large number of samples in both classes, owing to the low prevalence of AF in the general population \cite{4}. 
\\

Data augmentation not only can help mitigate overfitting when training supervised learning models \cite{11,12,13}, but also can increase the sample size by generating synthetic samples from real ones so that machine learning models can be developed based on a dataset of limited sample size. 
\begin{figure}[htbp]
  \centering
  \includegraphics[width=1\columnwidth]{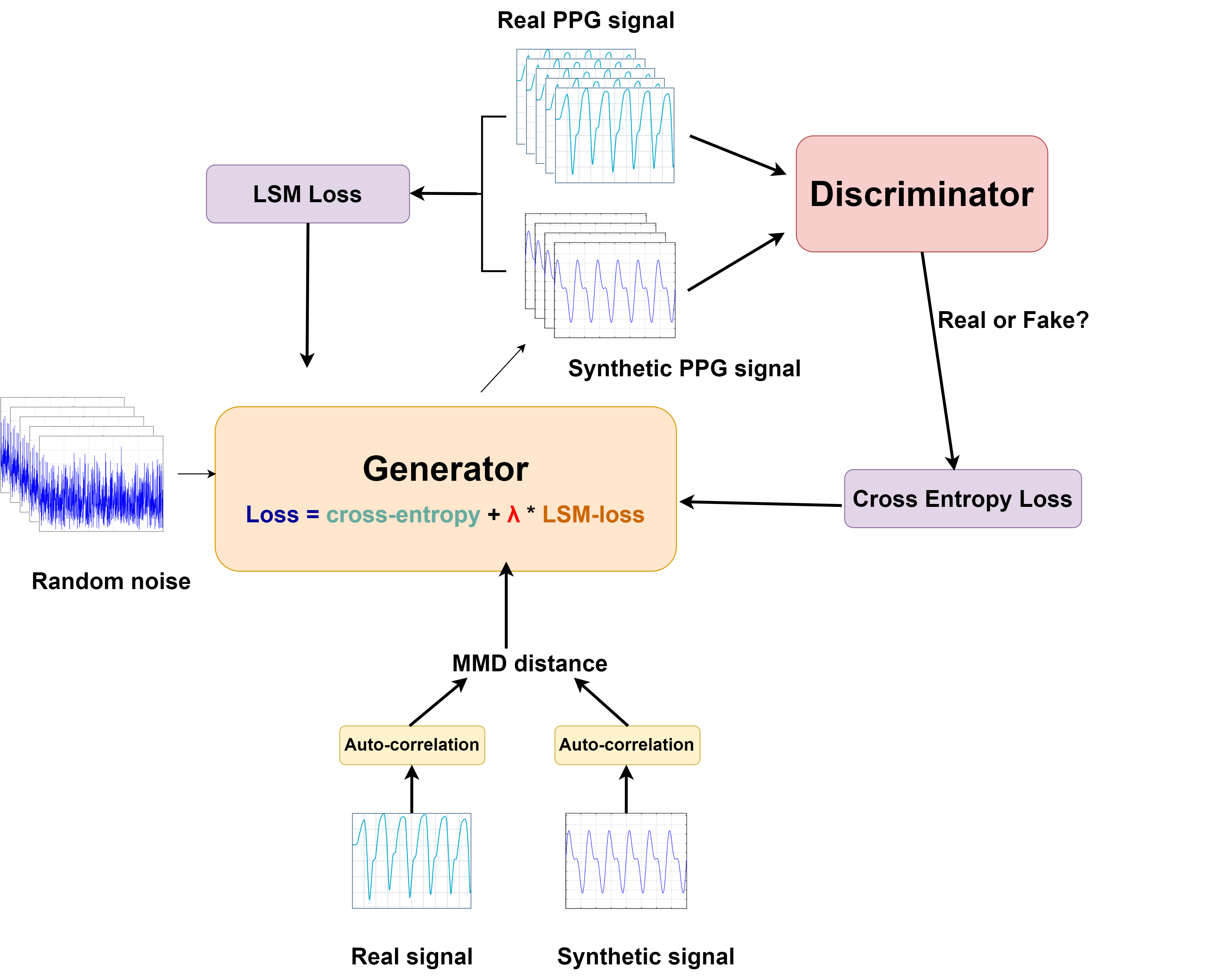}
  \caption{The overall workflow of LSM-GAN. LSM-loss is integrated with adversarial loss in a weighting paradigm. The weight parameter is selected by the shortest MMD distance between the autocorrelation of real and synthetic signals.}
\label{fig1}
\end{figure}
\\
To investigate suitable augmentation techniques for the PPG-based AF detection task, the present study started with various popular GAN constructs, including deep convolutional GAN (DCGAN) \cite{17}, Wasserstein DCGAN (W-DCGAN) \cite{18}. However, we quickly realized that these off-the-shelf techniques do not work well for PPG-based AF detection as they show a very limited amount of improvement over simple data augmentation techniques. Therefore, we propose a new GAN to generate synthetic PPG signals. In this new GAN, we adopted a different generator architecture as well as a new loss function that measures how close a synthetic signal is to a PPG in the frequency domain. The objective of this new GAN is to generate synthetic PPGs with a power spectrum close to the real ones. Therefore, we call this GAN Log-Spectral matching (LSM)-GAN and the new loss function LSM-loss, as shown in Figure \ref{fig1}. Our results show that the LSM-GAN generates synthetic AF signals with a distribution closest to the real ones and achieves the greatest performance gain in AF detection, compared to the other two GANs and two conventional data augmentation techniques. The main contributions of this paper include:

\begin{itemize}
\item To our knowledge, this is the first work that incorporates spectral information into the loss function for augmenting  PPG signals and considers using GAN to generate synthetic PPG data for the task of AF detection. 
\item A weighting mechanism is implemented to balance the LSM-loss and adversarial loss, and an algorithm to automatically search for the optimal weighting parameter is proposed.  
\item Besides testing on the internal dataset, we evaluate trained models from each augmentation method on two public PPG datasets, which validates the generalizability of the proposed LSM-GAN.
\end{itemize}

The rest of this paper is organized as follows: section \ref{sec2} describes related prior works on AF detection and physiological signal augmentation. Details of our experiments, including datasets and training procedures, as well as our proposed method, are provided in Section \ref{sec3}. Experiments are presented in Section \ref{sec4}, followed by a Result section \ref{sec5}, then discussion and summary of our work in Section \ref{sec6}.

\section{Related work}
\label{sec2}

Several studies have developed GANs to generate synthetic PPG signals. PlethAugment \cite{15}, implemented a more advanced technique, generative adversarial network (GAN) \cite{16}, for PPG data augmentation. Three different conditional GANs were tested on various public datasets for different classification tasks, showing that GAN can help generate realistic PPG signals and improve the performance of PPG-based models. This study also sheds light on the effect of synthetic data on class imbalance and the influence of different ratios of real-world to artificial training data on classification performance. However, the performance comparison of GANs with traditional augmentation techniques, such as shifting and cropping, was not conducted in the study, which is a missed opportunity to inform whether an optimal choice of data augmentation solutions is task-specific. Furthermore, the tasks investigated in the study did not include AF detection. \citet{23} introduced a GAN to generate high-quality PPG signal from the simultaneously recorded ECG. The architecture contains a Bi-LSTM based generator and 1D-CNN based discriminator. However, the proposed GAN can only take 1-second ECG and generate 1-second PPG at a time. One has to stitch consecutive one-second of PPGs in order to get a longer duration signal.  In SynSigGAN \cite{24},  a GAN model was proposed to generate four kinds of physiological signals (electrocardiogram (ECG), electroencephalogram (EEG), electromyography (EMG), PPG). In the preprocessing stage, each signal goes through a discrete wavelet transformation and an inverse discrete wavelet transformation, and the signal denoising process takes place in between. As the last part of preprocessing, automatic segmentation is applied to set the length for each type of physiological signal from GAN. Again, no downstream use case was investigated to evaluate the efficacy of the synthetic signals. \citet{25} proposed a cycleGAN \cite{26} based approach to generate PPG signal for respiratory rate (RR) estimation. A novel loss function was introduced, which takes the RR of synthetic signal into account. Results showed that, by adding the synthetic PPG signals, the accuracy of RR estimation outperformed other state-of-the-art methods using an identical experiment setting and dataset. The study suggests that introducing task-related information into the loss function is a promising solution to improve many GAN-based tasks.

\section{Methods}
\label{sec3}

\subsection{Objective and Proposed Architecture}

Integrating additional loss into the cost function of the generator has been proven to be helpful for generating synthetic biomedical signals \cite{24}. Also, in the speech signal area, introducing information from spectral-domain has brought significant improvement in speech recognition and speech waveform generation tasks \cite{27,28}. In the present study, we propose to include spectral information from PPG waveforms into the cost function for GAN. The hallmark of AF compared to normal sinus rhythm lies in the irregular irregularity in the rhythm. Therefore, synthetic signals that retain spectral characteristics of  real ones will likely improve the model performance. Additionally, because our AF detector and most reported deep neural network approaches process PPG signals in time domain, matching in spectral domain still allows randomness in the phases of synthetic signals and hence enriches training data in a profound way. We hypothesize that such randomness will be beneficial for training AF detectors.
\\

\begin{figure}[htbp]
\centering
\includegraphics[width=1\columnwidth]{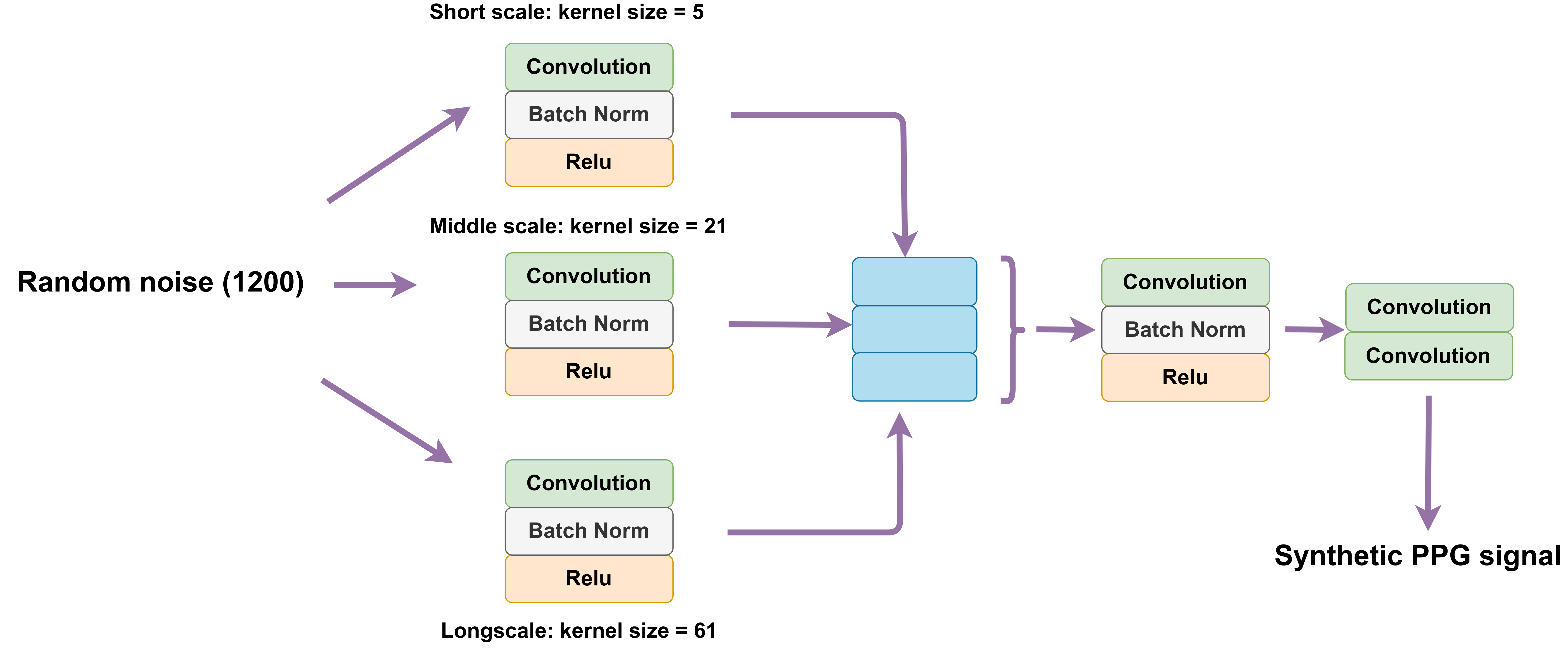}
\caption{The network architecture: discriminator and generators with conventional and new architectures.}
\label{fig2}
\end{figure}

GAN is composed of two neural networks: \textbf{generator(G)} and \textbf{discriminator(D)}, the generator takes random noise as input to generate synthetic signals. However, the length of the input noise is an arbitrary choice and few studies have investigated its influence. In the present study, instead of following a conventional choice - 100 - as the dimension of random noise, we choose an identical length to the output dimension as the input (1200 in this study). To achieve this, we build a new architecture for the generator which is able to generate the synthetic signal with an identical length as input. The comparison of two different generators is shown in Figure \ref{fig2}. To better distinguish GANs with different generator architectures, we use \textbf{DCGAN-100} and \textbf{DCGAN-1200} to name two baseline methods with different input dimensions in later sections.
\subsection{LSM-Loss}

\begin{figure}[ht]
\centering
\subfigure[Matching distance]{
\begin{minipage}[t]{\linewidth}
\centering
\includegraphics[width=0.9\columnwidth]{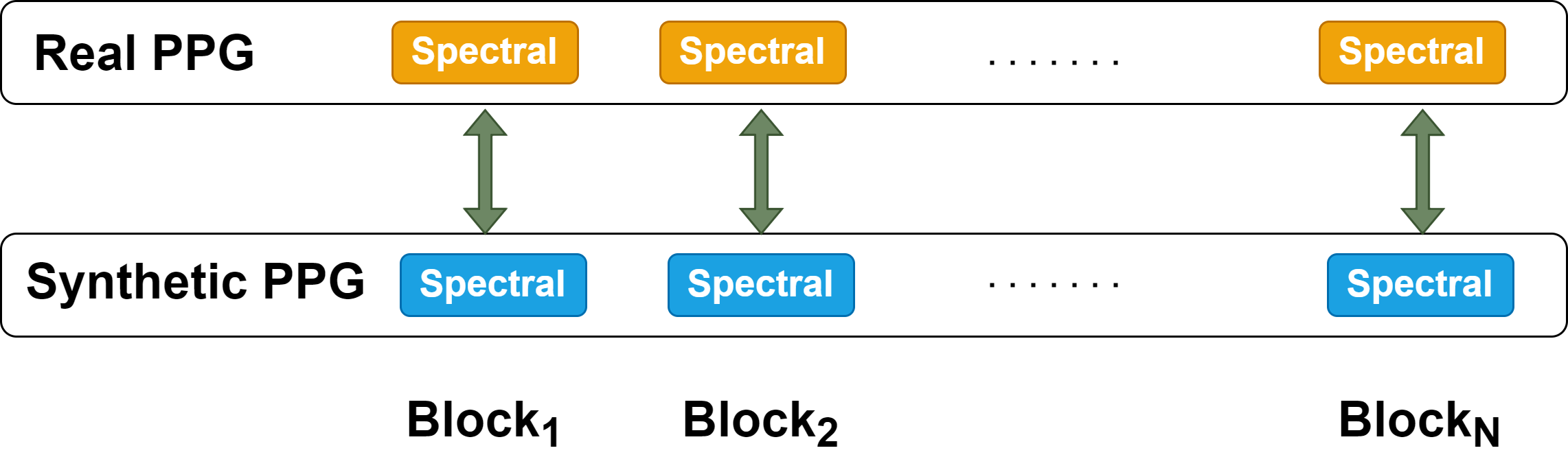}
\end{minipage}}\\
\subfigure[Self-consistency distance]{
\begin{minipage}[t]{\linewidth}
\centering
\includegraphics[width=0.9\columnwidth]{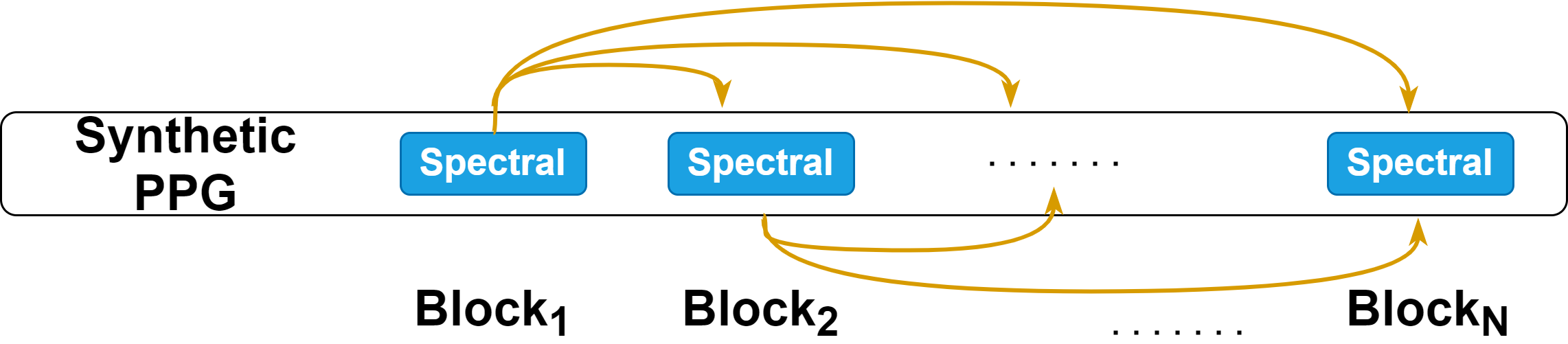}
\end{minipage}}
\caption{The proposed two distances: matching distance ($d_m$) and self-consistency distance ($d_s$). Matching distance is designed to measure the difference between blocks in real and synthetic signals with same index. Self-consistency distance measures the difference between blocks within one synthetic signal.}
\label{fig3}
\end{figure}

We divide a PPG strip into successive blocks and calculate the periodogram of each block separately. We define two distances: matching distance ($d_m$) and self-consistency distance ($d_s$).  Matching distance is designed to measure the difference between matching blocks in real and synthetic signals. Self-consistency distance measures the difference among blocks within one synthetic signal. These distance metrics are defined as: 
\begin{equation}
\label{eq1}
d_{m_{i}}=\left\|p s d\left(b_{i}\right)-p s d\left(\widehat{b}_{i}\right)\right\|^{2}, 
\end{equation}
\begin{equation}
\label{eq2}
d_{s_{i, j}}=\left\|p s d\left(\hat{b}_{i}\right)-p s d\left(\widehat{b}_{j}\right)\right\|^{2}, i<j
\end{equation}
where $\widehat{b}_{i}$ and $b_j$ represent $i$-th block from synthetic and real PPG, respectively. $\widehat{b}_{i}$ and $\widehat{b}_{j}$ are the two different blocks within one synthetic PPG, $i,j\in \{1,2,3,\cdots , N\}$ and $N$ is the number of blocks for one PPG. $psd(\cdot)$ is the function that returns the magnitude of spectrum for the input time sequence. As illustrated in Figure \ref{fig3}, the matching distance is calculated by the L2 norm between the spectra of a matched real and a synthetic PPG signal block. Self-consistency distance is calculated by measuring the L2 norm among different blocks of a synthetic PPG signal itself. 

LSM-loss also considers aggregating $d_m$ and $d_s$ for blocks in one PPG through a weighted paradigm. However, directly averaging may not be the best approach to aggregate each block. For example, irregular pulses can exist anywhere in one AF-PPG segment, and the spectral value of blocks having irregular pulses would be significantly different from other blocks. But averaging all blocks will dilute the contribution from those irregular pulses. Such a situation will not be optimal for matching spectral patterns between a real and a synthetic AF signal. We hence utilize the aggregation function $F\in \{Mean,Max\}$ as a hyperparameter to accommodate potentially different needs in generating  AF and  Non-AF signals.

The adversarial loss is defined as,
\begin{equation}
\label{eq3}
L_{a d v}(G, D)\!=\! E_{z \sim N(0,1)}, x \!\sim\! p(data)\!\left[(1\!-\! D(G(z)))^{2}\right]\!,
\end{equation}
where $z$ represents the input noise, and $G(z)$ is the synthetic signal. The final LSM-loss is defined as the linear combination of three losses described above,
\begin{equation}
\label{eq4}
\begin{split}
L_{l s m}=~&L_{a d v}(G, D)+\lambda_{1} \times F\left(\left\{d_{m_{1}}, \ldots, d_{m_{N}}\right\}\right)\\
&+\lambda_{2} \times F\left(\left\{d_{s_{1,2}}, d_{s_{1,3}}, \ldots\right\}\right),
\end{split}
\end{equation}
where $N$ represents number of blocks for one PPG,  $\lambda_{1}$ and $\lambda_{2}$ are the hyperparameters balancing the three losses, and $F$ is either Mean or Max function.

\subsection{Hyperparameters selection}

There are three hyperparameters in equation \eqref{eq4} that need to be optimized including $\lambda_{1}$, $\lambda_{2}$ and $F$. We design the loss function in this flexible way because we anticipate the different choices of hyperparameters in this loss function would be needed to accommodate training GANs to generate AF vs Non-AF signals. For an AF PPG, self-consistency loss would be less important but the matching of spectra at a block level would be critical. On the other hand, for a non-AF PPG (most of which correspond to sinus rhythm), self-consistency would be needed to ensure a more realistic artificial signal.

\begin{algorithm}[htbp]
   \caption{Hyperparameter selection process for LSM-GAN. We select $k=300$ samples from each signal set. \emph{\textbf{Least distance} = infinity, \textbf{Best set} = \{\}}}
   \label{alg1}
\begin{algorithmic}
   \FOR{$\lambda_{1}$ in $[0,3]$ with step size $0.1$}
   \FOR{$\lambda_{2}$ in $[0,3]$ with step size $0.1$}
   \FOR{$F$ in  {\bfseries \{Mean, Max\}}}
   \STATE 1. Sample $k$ signal $\{x^1,\cdots,x^k\}$ from data generating distribution $p_{data}(x)$
   \STATE 2. Sample $k$ noise samples $\{z^1,\cdots,z^k\}$ from noise prior $p_g(z)$
   \STATE 3. Generate $\{\hat{x}^{1},\cdots,\hat{x}^{k}\}$ synthetic signals with trained LSM-GAN $G(z)$ with $\{\lambda_{1},\lambda_{2},F\}$
   \STATE 4. Calculate the MMD distance:
   \STATE \quad \emph{\textbf{current}} $=\frac{1}{k} \!\sum\limits_{i=1}^{k}\! M\! M\! D\!\left[autocorrelation\!\left({x}^{i}\right)\!,\right.$ 
   \STATE \quad $\left.autocorrelation \left(\widehat{{x}}^{i}\right)\right]$
   \STATE 5. \emph{\textbf{Best set}} $=\{\lambda_{1},\lambda_{2},F\} $ \textbf{if \emph{current $<$ Least distance}}
   \ENDFOR
   \ENDFOR
   \ENDFOR
   \STATE Return the \emph{\textbf{Best set}}.
\end{algorithmic}
\end{algorithm}

Different from many other studies which set the weights manually, we select the three hyper-parameters $\lambda_{1}$, $\lambda_{2}$ and $F$ by optimizing a guided grid search process, as shown in algorithm \ref{alg1}. First, we choose a large range of $[0, 3]$ for $\lambda_{1}$ and $\lambda_{2}$ with a step size of 0.1, and \{Mean, Max\} for $F$. Second, for each combination of $[\lambda_{1}, \lambda_{2}, F]$, a set of 300 synthetic PPG signals from LSM-GAN will be generated. Another set of 300 real signals will be randomly selected from the training set. Third, autocorrelations are calculated for both real and synthetic signals, then the maximum mean discrepancy (MMD) distances of autocorrelations between the real signal set and each of the synthetic signal sets are calculated. Lastly, the best combination of $[\lambda_{1}, \lambda_{2}, F]$ will be chosen based on the shortest MMD distance between synthetic signals and real signals.

\subsection{Baseline data augmentation methods}

\textbf{Data-copying:} As the most straightforward data augmentation method, data-copying simply duplicates randomly selected signals and then adds them into the training set.

\noindent\textbf{Permutation:} A 30-second PPG signal is divided into five equal-length sub-segments, then all five sub-segments are rearranged by randomly permuting their orders and concatenated to form a new 30-second signal.

\noindent\textbf{Deep convolutional generative adversarial network (DCGAN) and DCGAN with Wasserstein distance (W-DCGAN):}

GAN has two components \cite{16}: a generator network $G$ and a discriminator network $D$, as shown in Figure \ref{fig1}. $G$ receives a random signal $z$ and generates a `fake' PPG signal. $D$ is a binary classifier to determine whether the input signal is a real or fake PPG. The training process of GAN is a zero-sum game, where generator and discriminator aim to minimize the objective function:
\begin{equation}
\begin{split}
\min _{G} \max _{D} V(D, G)=~&E_{x \sim p_{\text {data }(x)}}[\log D(x)]\\
&+E_{z \sim p_{z(z)}}[\log (1-D(G(z)))]
\end{split}
\end{equation}
where $x$ is the real PPG signal and $G(z)$ is the synthetic data generated from random signal $z$. $D(x)$ and $D(G(z))$ are the discriminator's estimates of the probability of $x$ and $G(z)$ being real, respectively.

Convolutional neural network (CNN) based models are capable of learning good feature representations of the input data, leading to state-of-the-art performance in many classification tasks. Compared to vanilla GAN, DCGAN  adopts the structure of deep CNN to improve the quality of generated signal and accelerate the converging process.

The Wasserstein deep convolutional generative adversarial network (W-DCGAN) is a further extension of DCGAN. In DCGAN, we only change the model structure and keep the same cost function as vanilla GAN. By introducing the Wasserstein distance, W-DCGAN not only improves the training stability but also has a cost function that is related to the quality of the generated signal. The new cost function is 
\begin{equation}
\min _{G} \max _{D} L_{W G A N}(D, G)=-E_{x \sim p_{\text {data }(x)}}[D(x)]+E_{z \sim p_{z(z)}}[D(G(z))]
\end{equation}

\begin{table*}[htbp]
\caption{The number of records in the training and test sets.}
\label{tab1}
\vskip 0.15in
\begin{center}
\begin{small}
\begin{sc}
\begin{tabular}{ccccc}
\toprule
& \multicolumn{2}{c}{Training set} & \multicolumn{2}{c}{Test set}\\
\midrule
Center & \multicolumn{2}{c}{UCLA medical center} & \multicolumn{2}{c}{UCSF Neuro ICU}\\
Number of patients & \multicolumn{2}{c}{126} & \multicolumn{2}{c}{13}\\
Age & \multicolumn{2}{c}{18 to 95 years (median 63)} & \multicolumn{2}{c}{19 to 91 (median $=$ 73.5)}\\
\multirow{2}{*}{Number of records} & AF & Non-AF & AF & Non-AF\\
& 36855 & 249278 & 1216 & 1467\\
Total & \multicolumn{2}{c}{176133} & \multicolumn{2}{c}{2683}\\
Percentage & 15.24\% & 84.75\% & 45.32\% & 54.68\%\\
\bottomrule
\end{tabular}
\end{sc}
\end{small}
\end{center}
\vskip -0.1in
\end{table*}

\section{Experiments}
\label{sec4}

\subsection{Datasets}

\textbf{Training data:} Continuous fingertip PPG (fPPG) recordings were collected with pulse oximeters from 126 in-hospital patients aged between 18 and 95 years (median 63) who were admitted to UCLA Medical Center between April 2010 and March 2013. A board-certified cardiac electrophysiologist marked the start and end of an AF episode based on co-registered ECG recordings. Continuous PPG recordings were divided into consecutive non-overlapping 30-second records. Each record was labeled ``AF" or ``non-AF" depending on if it was extracted within or outside an AF marked episode, respectively.

\textbf{Testing data:} A set of continuous PPG data were collected from wearable devices (Empatica E4) worn by 13 acute stroke patients admitted into the neurological intensive care unit (NICU) of UCSF Medical Center between October 2016 and January 2018. Patients' age was between 19 to 91 (median $=$ 73.5). With the same method used in the training data, the continuous signals in the test set were segmented into consecutive non-overlapping 30-second segments (2683 in total). Table \ref{tab1}  shows the distribution of AF and Non-AF segments in the training and testing sets.

\begin{table*}[t]
\caption{The performance of AF detection from each augmentation method.}
\label{tab2}
\begin{tabular}{cccccc}
\toprule
 & Accuracy & Sensitivity & Specificity & PPV & NPV\\
\midrule
Original & 0.8210 & 0.6060 & \textbf{0.9993} & \textbf{0.9986} & 0.7537\\
Data-Copying & 0.9288 & 0.8593 & 0.9863 & 0.9812 & 0.8943\\
Permutation & 0.9369 & 0.8848 & 0.9850 & 0.9799 & 0.9117\\
DCGAN-100 & 0.9213 & 0.8470 & 0.9829 & 0.9763 & 0.8857\\
DCGAN-1200 & 0.9284 & 0.8684 & 0.9781 & 0.9705 & 0.8996\\
W-DCGAN-100 & 0.9250 & 0.8511 & 0.9863 & 0.9810 & 0.8888\\
W-DCGAN-1200 & 0.9370 & 0.8799 & 0.9843 & 0.9789 & 0.9081\\
LSM-GAN & \textbf{0.9612} & \textbf{0.9284} & 0.9884 & 0.9851 & \textbf{0.9433}\\
\bottomrule
\end{tabular}
\end{table*}

\subsection{Experiment design}

Three experiments are designed to evaluate the proposed LSM-GAN against baseline methods.  Each experiment tests a specific aspect of practical relevance when considering a GAN-based data augmentation strategy, including inter-class balancing, resilience to artifacts and the training sample size. Resnet-50 is used here as the classifier for all the experiments.

\textbf{Experiment 1:} In our study, the ratio between AF sample size and that of Non-AF is approximately $1:7$. To investigate whether a more balanced training data would help improve the final classification accuracy, we augmented only the AF cases by 6 folds to achieve the inter-class balance in the first experiment. The proposed LSM-GAN and baseline models are all trained based on the same set of AF data in the training set and then the resultant models are used to generate 212,423 synthetic AF-PPG strips per each model to augment the training data. Various performance metrics, including accuracy, sensitivity, specificity, positive prediction value (PPV) and negative prediction value (NPV), are used to compare the classifiers that are trained with the augmented dataset from different data augmentation approaches.

\begin{table*}[t]
\caption{AF detection performance under different hyperparameters for AF signals.}
\label{tab4}
\begin{tabular}{cccccc}
\toprule
 & Accuracy & Sensitivity & Specificity & PPV & NPV\\
\midrule
$\lambda_{1} = \lambda_{2} = 1$, $F = Mean$ (no weight mechanism) & 0.9467 & 0.9095 & 0.9775 & 0.9710 & 0.9287\\
$\lambda_{1}=2.2$, $\lambda_{2} = 0.6$, $F=Max$ (largest MMD) & 0.9306 & 0.8651 & 0.9850 & 0.9795 & 0.8980\\
$\lambda_{1}=\lambda_{2} = 1.5$, $F = Mean$ (optimized hyperparameters) & \textbf{0.9612} & \textbf{0.9284} & \textbf{0.9884} & \textbf{0.9851} & \textbf{0.9433}\\
\bottomrule
\end{tabular}
\end{table*}

\textbf{Experiment 2:} The issue of signal quality cannot be overlooked for PPG-based studies, which is especially true in ambulatory settings. Therefore, we dedicate this experiment to evaluating the performance of different augmentation techniques at various levels of artifacts by investigating the relationship between F1 scores and the proportion of artifacts within the PPG signals. We split the testing set into four groups based on the percentage of artifacts: clean (0\%), (0\% - 20\%), [21\%-60\%) and [61\%-100\%]. Then we pick models from experiment 1 for each augmentation method and test them on those four groups separately. 

\textbf{Experiment 3:} To test the effect of training sample size, we constructed a series of balanced training sets with an increasing sample size from 300,000 to 2,000,000. For each training set, if the number of required cases (e.g., 15,000 AF cases are needed for the total sample size of 30,000) is less than the original cases, then they are randomly selected within the existing data. When the required case exceeds the original size, additional samples will be generated by different augmentation methods for both AF and non-AF cases. All the trained models are tested on the same four signal quality groups as in experiment 2.

\subsection{Preprocessing and Training}

Raw PPG signals collected in this study were sampled at 240 Hz, we first down-sampled the PPG signals to a sampling rate of 40 Hz. Then, we applied a band-pass FIR filter with a pass-band frequency of 0.9 Hz and stop-band frequency of 5 Hz on the PPG signals. Finally, the min-max normalization was performed on PPG segments to ensure all signals are in the same scale. The proposed LSM-GAN and other GANs were trained from scratch on AF signals and Non-AF signal, separately. We used 200 as the batch size and then trained the GANs for epochs at a learning rate of 0.001. The learning rate decayed 0.0001 for each epoch. Early stopping was performed when the loss on validation did not improve in 6 epochs.The Adam optimization algorithm and cross-entropy loss function were used to train the ResNet-50 with 512 being the mini-batch size and at a learning rate of 0.001. To avoid overfitting, we also employed an early stopping procedure which stops the training procedure if the validation loss does not improve in 6 epochs. The GANs were implemented using Pytorch and the Resnet-50 was implemented in Keras using TensorFlow backend, and the experiments were performed on a workstation with one NVIDIA RTX 1080Ti GPU and 64 GB memory. 

\section{Results}
\label{sec5}

\subsection{Experiment 1: Performance comparison between data augmentation methods}

Table \ref{tab2} summarizes the performance of AF detection for different combinations of original and augmented data sets. Compared to training with the original dataset, a balanced training set by simply duplicating all the AF cases 6 times through data-copying would increase 11\% of accuracy and 25\% of sensitivity. Permutation, DCGAN and W-DCGAN achieve similar performance gain to data-copying, while the proposed LSM-GAN offers the most performance improvement, with a 24\% gain in accuracy and 32\% in sensitivity with around 1\% reductions in specificity and PPV. We notice that traditional data augmentation methods and basic GAN models with conventional generator achieve similar performance, which all arrive at an accuracy of around 93\%. With the new architecture design, DCGAN-1200 and WDCGAN-1200 both perform better than the ones with the conventional generator. Built on top of the new architecture of the generator, the proposed LSM-GAN integrates the LSM-loss component offers an additional 3\% improvement in accuracy and 5\% in sensitivity.

\begin{table}[htbp]
\caption{The selected optimal hyper-parameters.}
\label{tab3}
\begin{tabular}{cccc}
\toprule
 & $\lambda_{1}$ & $\lambda_{2}$ & $F$ \\
\midrule
AF & 1.5 & 1.5 & Mean\\
Non-AF & 0.8 & 3.0 & Max\\
\bottomrule
\end{tabular}
\end{table}

Table \ref{tab4} reports the performance for classifiers trained on augmented data generated from LSM-GAN under different hyperparameters. Two different hyperparameter set are investigated, $\{1,1,Mean\}$ and $\{2.2,0.6,Max\}$. $\{1,1,Mean\}$ represents no weight mechanism which three loss terms are treated equally, and $\{2.2,0.6,Max\}$ is selected by the largest MMD distance calculated according to Algorithm \ref{alg1}. Results show that synthetic signals generated from LSM-GAN with optimized hyperparameters will help increase 1.5\% accuracy compared to LSM-GAN without weight mechanism and help increase 3\% in accuracy compared to LSM-GAN with the worst hyperparameter set.

Table \ref{tab3} summarizes the hyper-parameters selected by Algorithm 1. For AF signal, match-loss and self-consistency loss share the same weight of 1.5. This result indicates that two losses play a similar role  when generating realistic AF signals. For Non-AF signals, self-consistency loss weights more than the other two losses, which is expected since the key characteristic of non-AF (Sinus rhythm most of the time) signals is periodicity. And the weight of 0.8 indicates the matching-loss is less important than the other two loss terms. In terms of $F$, results show that selecting the block with maximum loss value can help generate more realistic non-AF signals, while taking the average of all blocks is better for generating AF signals. 

\subsection{Experiment 2: Resilience to artifacts}

\begin{figure}[ht]
\centering
\includegraphics[width=1\columnwidth]{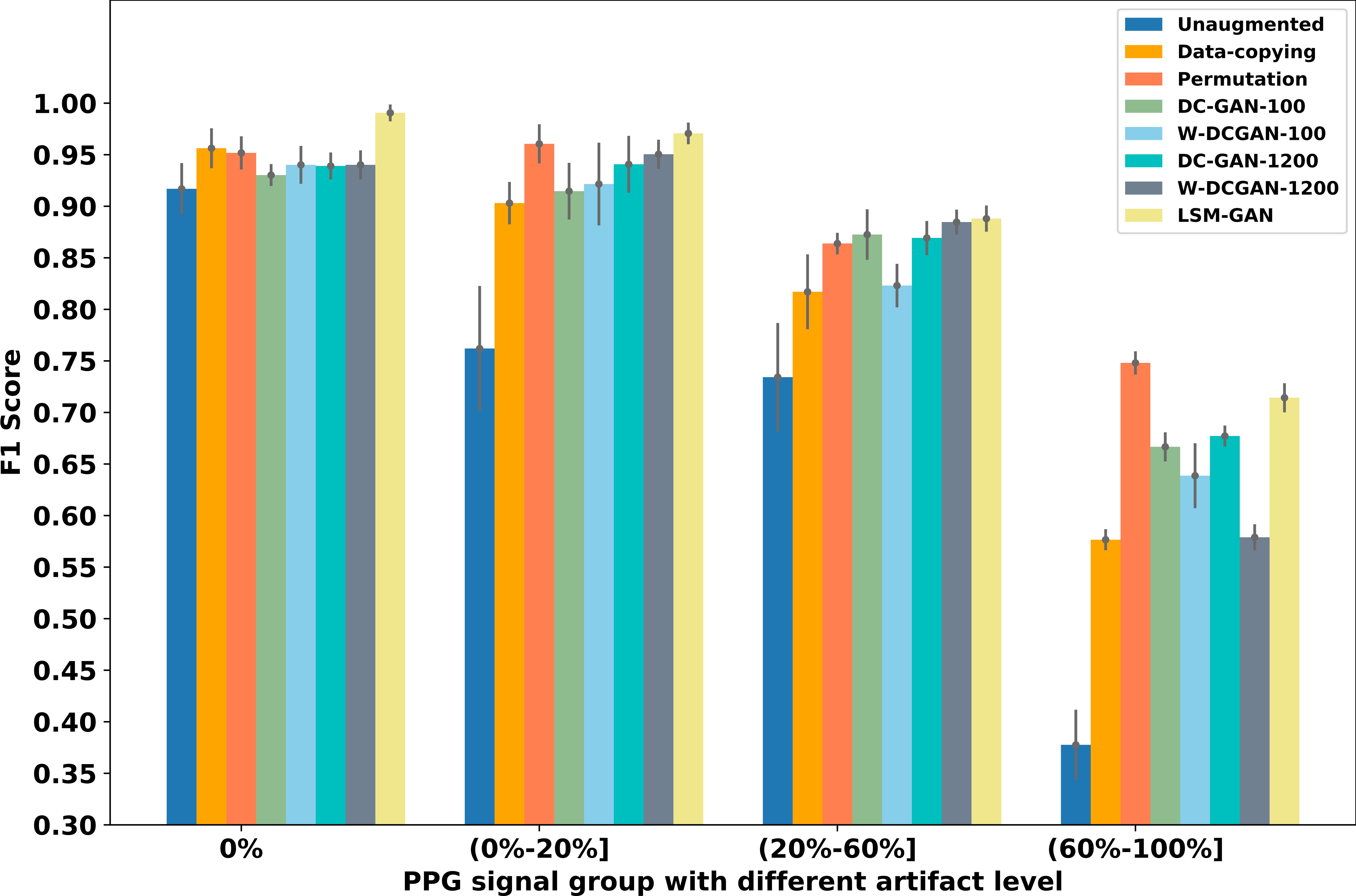}
\caption{Performance tested on PPG records with different percentage artifact level.}
\label{fig6}
\end{figure}

Figure \ref{fig6} compares each method's performance across the presence of different proportions of artifacts in PPG signals. F1 score is used because AF and Non-AF classes become unbalanced within each signal quality group.

It can be seen that all methods perform poorly in the poor-quality group (more than 60\% of artifacts), although substantial performance improvement can still be observed by different augmentation methods compared to the original dataset. On the other hand, all methods can achieve over 90\% F1 score for the excellent quality group (0\% of artifacts), especially for LSM-GAN which achieves 99\% F1 score (Sensitivity: 0.98\%; Specificity: 0.99\%; PPV: 0.99\%; NPV: 0.98\% ). 

We can observe that F1 score decreases with the increasing artifacts portion in PPG, except Permutation, it achieves a F1 score of 96\% in signal group (0-20\%] which is better than 95\% in the perfect signal quality group. Also, model trained on the original dataset has the least performance at each signal quality group, with F1 scores of 91\%, 76\%, 73\% and 38\%. At the same time, LSM-GAN achieves 99\%, 96\%, 90\% and 72\% separately in each signal quality group, which improves 8.7\%, 26\%, 23\% and 89\%, respectively. Other baseline methods also show improvement but not as significantly as LSM-GAN. However, Permutation achieves 75\% of F1 score in the signal quality group (60-100\%], which is higher than 73\% obtained by LSM-GAN. Also, Permutation has better F1 score compared to baseline GANs except in the signal quality group (20\%-60\%].

\subsection{Experiment 3: Data augmentation at different sample sizes}

To evaluate the effect of training sample size on the model performance, experiment 3 is conducted. A series of training sets with an increasing sample size from 300,000 to 2,000,000 is constituted. The same Resnet-50 is trained based on each training set separately and tested on the same four signal quality groups as experiment 2.  This process is repeated for all augmentation methods. 

\begin{figure*}[!t]
\centering
\subfigure[Signal quality group - 0\%]{
\begin{minipage}[t]{0.24\linewidth}
\centering
\includegraphics[width=1\columnwidth]{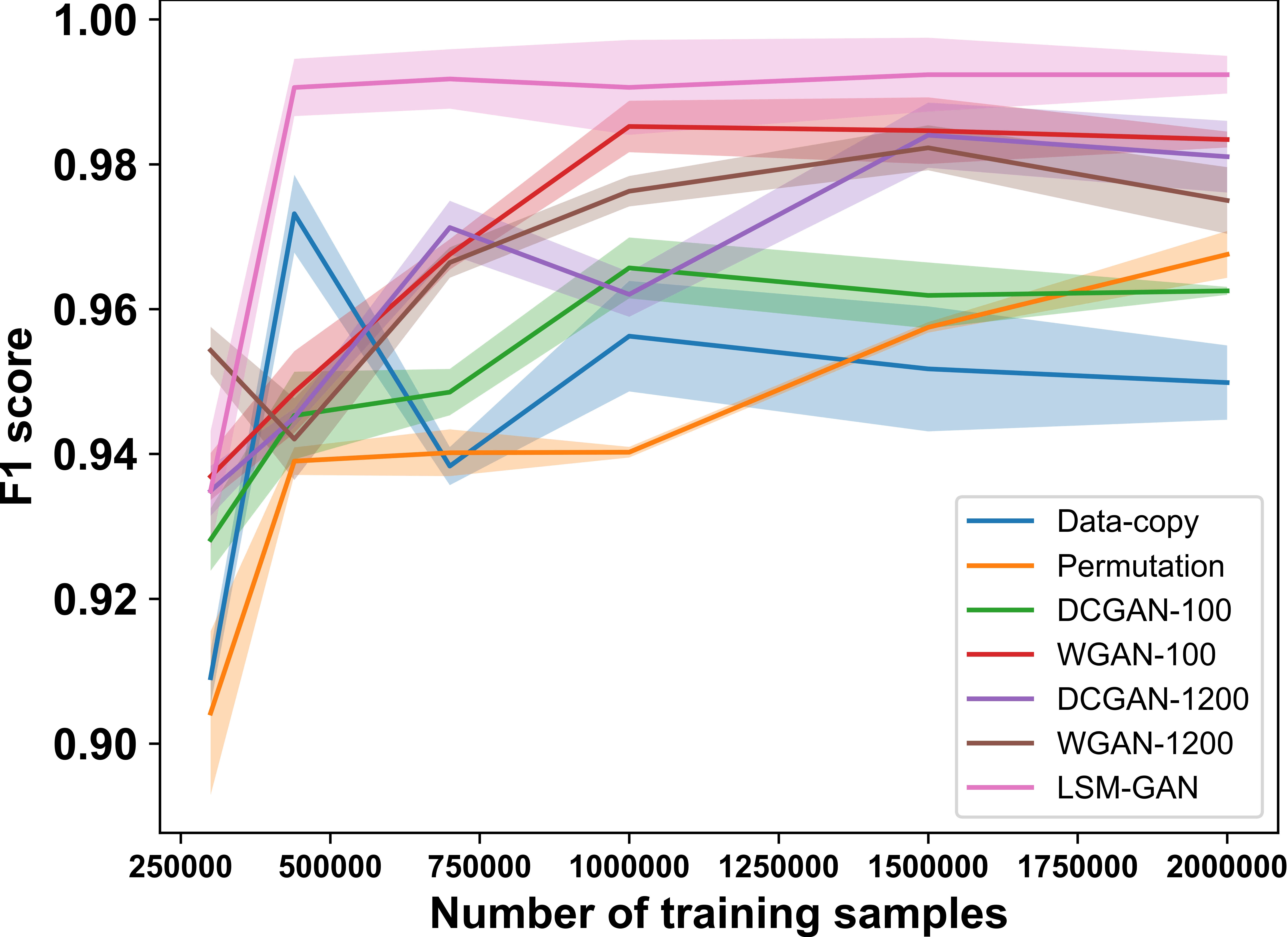}
\end{minipage}}
\subfigure[Signal quality group - (0\% -20\%$\text{]}$]{
\begin{minipage}[t]{0.24\linewidth}
\centering
\includegraphics[width=1\columnwidth]{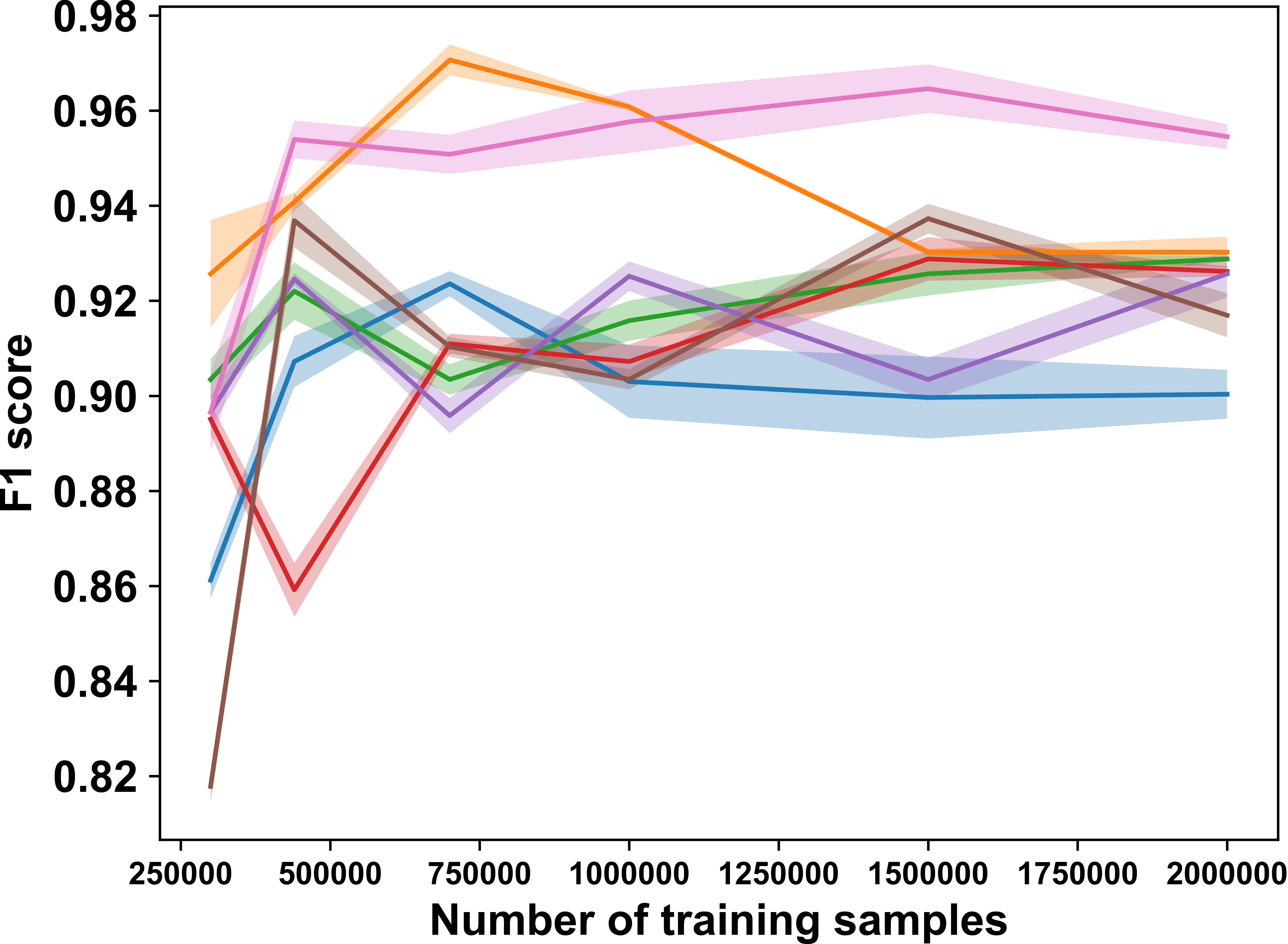}
\end{minipage}}
\subfigure[Signal quality group - (20\% -60\%$\text{]}$]{
\begin{minipage}[t]{0.24\linewidth}
\centering
\includegraphics[width=1\columnwidth]{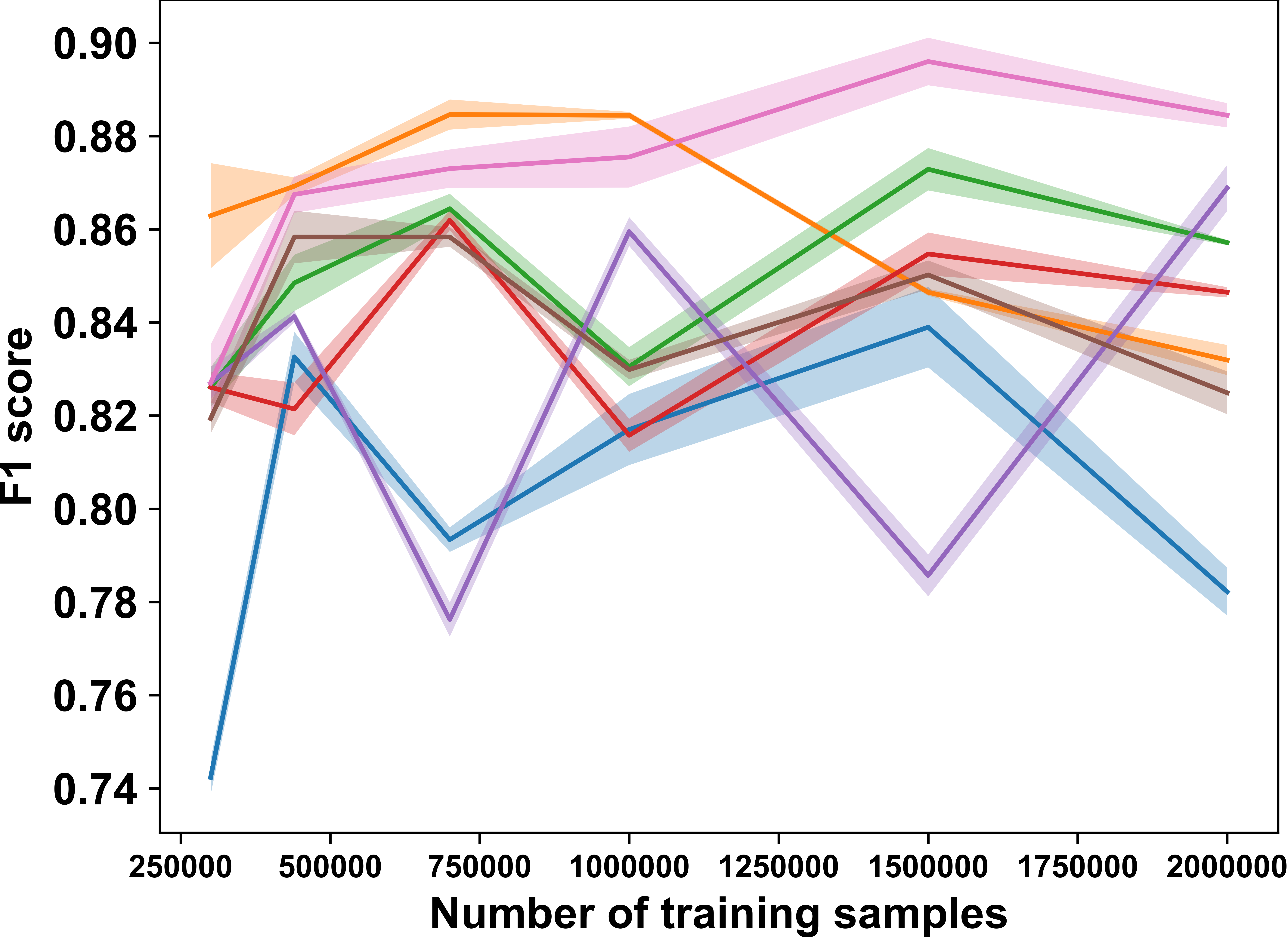}
\end{minipage}}
\subfigure[Signal quality group - (60\% -100\%$\text{]}$]{
\begin{minipage}[t]{0.24\linewidth}
\centering
\includegraphics[width=1\columnwidth]{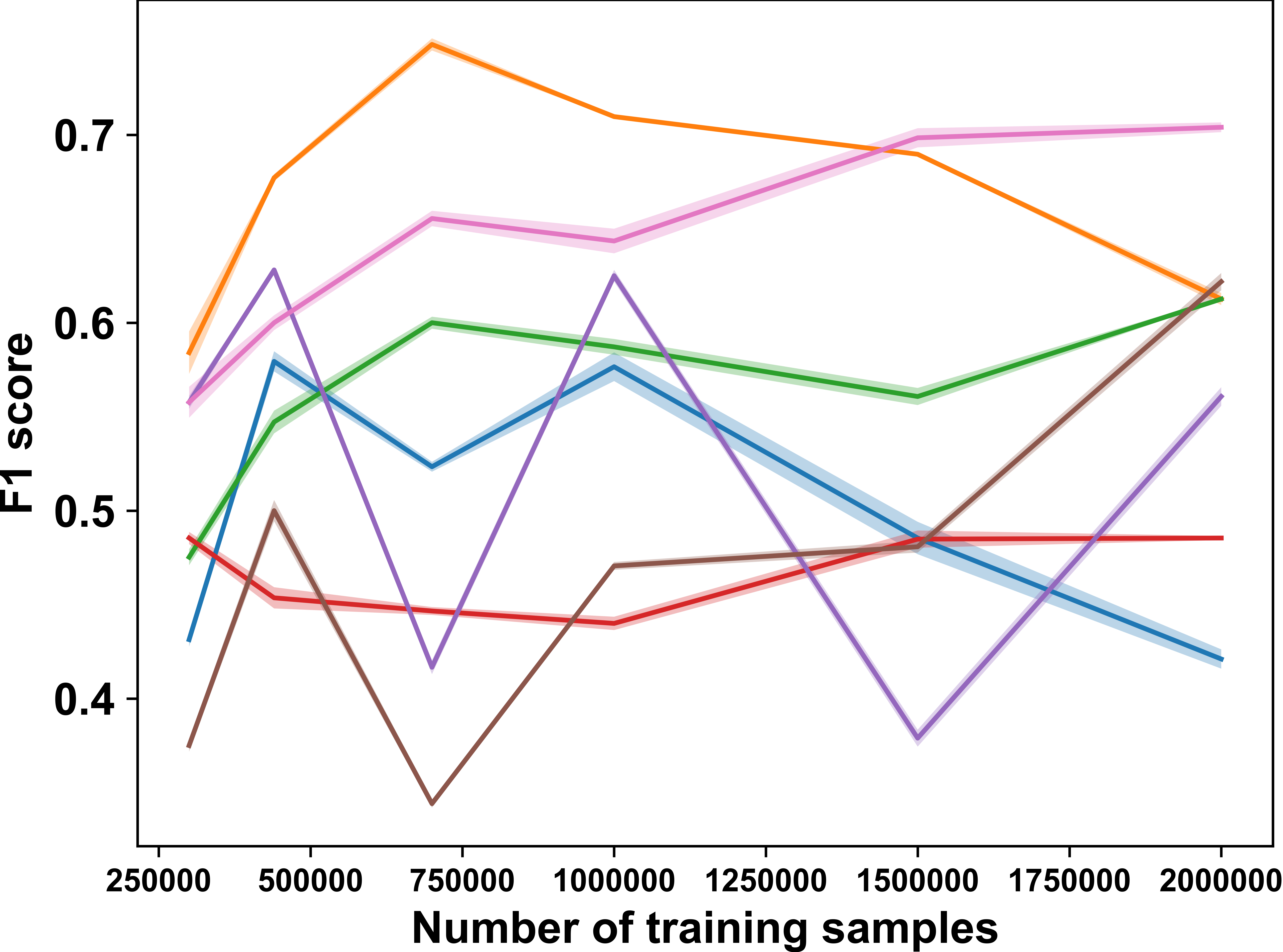}
\end{minipage}}
\caption{Comparison of F1 score for different training sample sizes on different signal quality groups. Four subplots have different scales for $Y$-axis in order to demonstrate the results in a better resolution.}
\label{fig7}
\end{figure*}

Figure \ref{fig7} compares the F1 score from models trained with data of different sample sizes that are generated with different data augmentation methods. We can observe that in Figure \ref{fig7} (a) when the signal quality is perfect, all the methods have a clear increasing trend with the added number of training samples except Data-copying. However, when the signal quality gets worse, in Figure \ref{fig7} (b), only WGAN-1200 and LSM-GAN still keep the increasing trend. Moreover, only LSM-GAN shows the steady uptrend in Figure \ref{fig7} c and d, where the signal quality is even worse and baseline methods have no positive relationship with added sample size. Another interesting observation is Permutation (orange curve). In Figure \ref{fig7} (a), Permutation has the least F1 score most of the time. While in Figure \ref{fig7} (b)-(d), when there are artifacts in PPG signals, Permutation shows great resilience to artifacts and keeps the leading performance along with LSM-GAN.

\subsection{External validation}

\begin{figure}[ht]
\centering
\subfigure[DeepBeat dataset]{
\begin{minipage}[t]{0.48\linewidth}
\centering
\includegraphics[width=1\columnwidth]{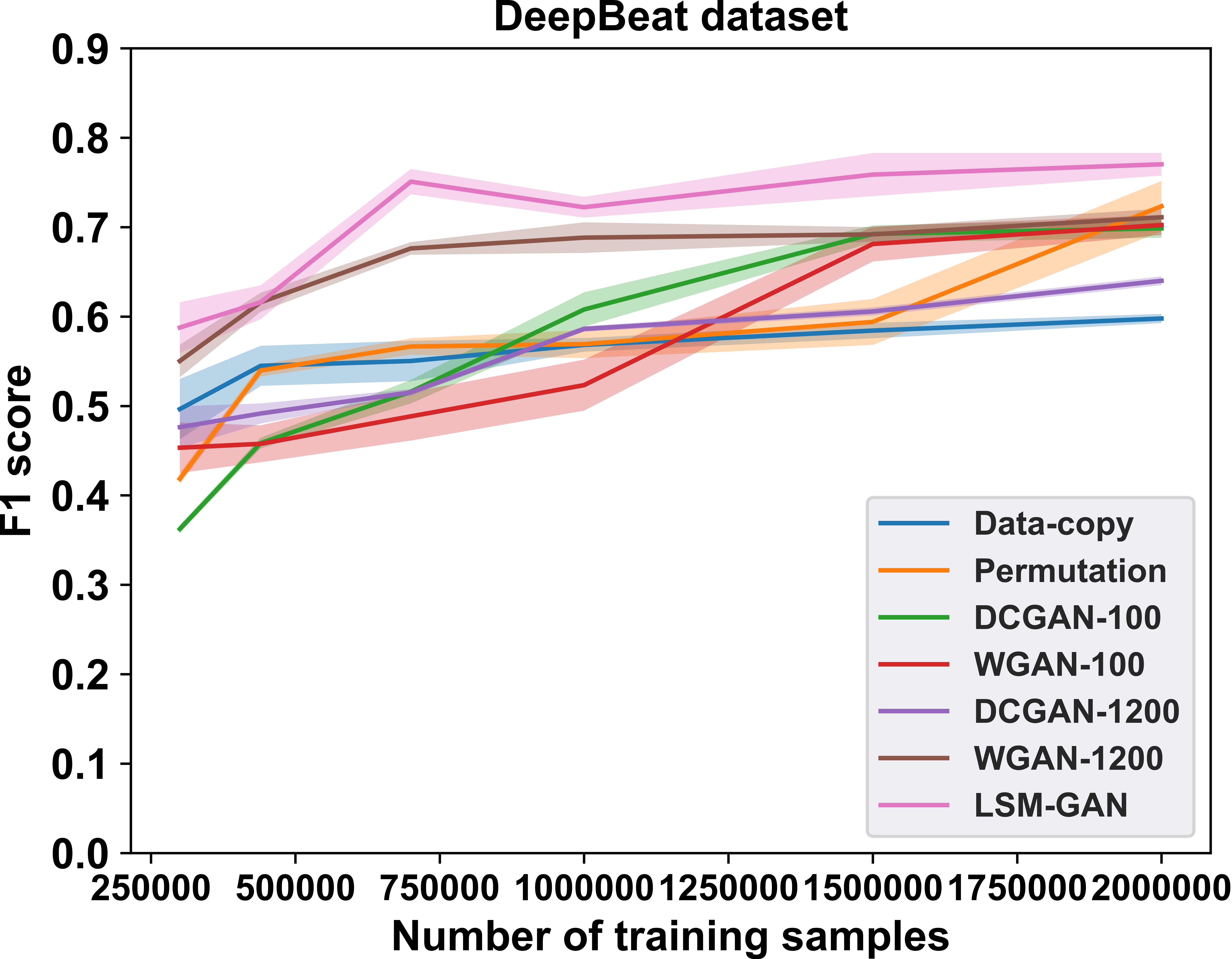}
\end{minipage}}
\subfigure[MIMIC dataset]{
\begin{minipage}[t]{0.48\linewidth}
\centering
\includegraphics[width=1\columnwidth]{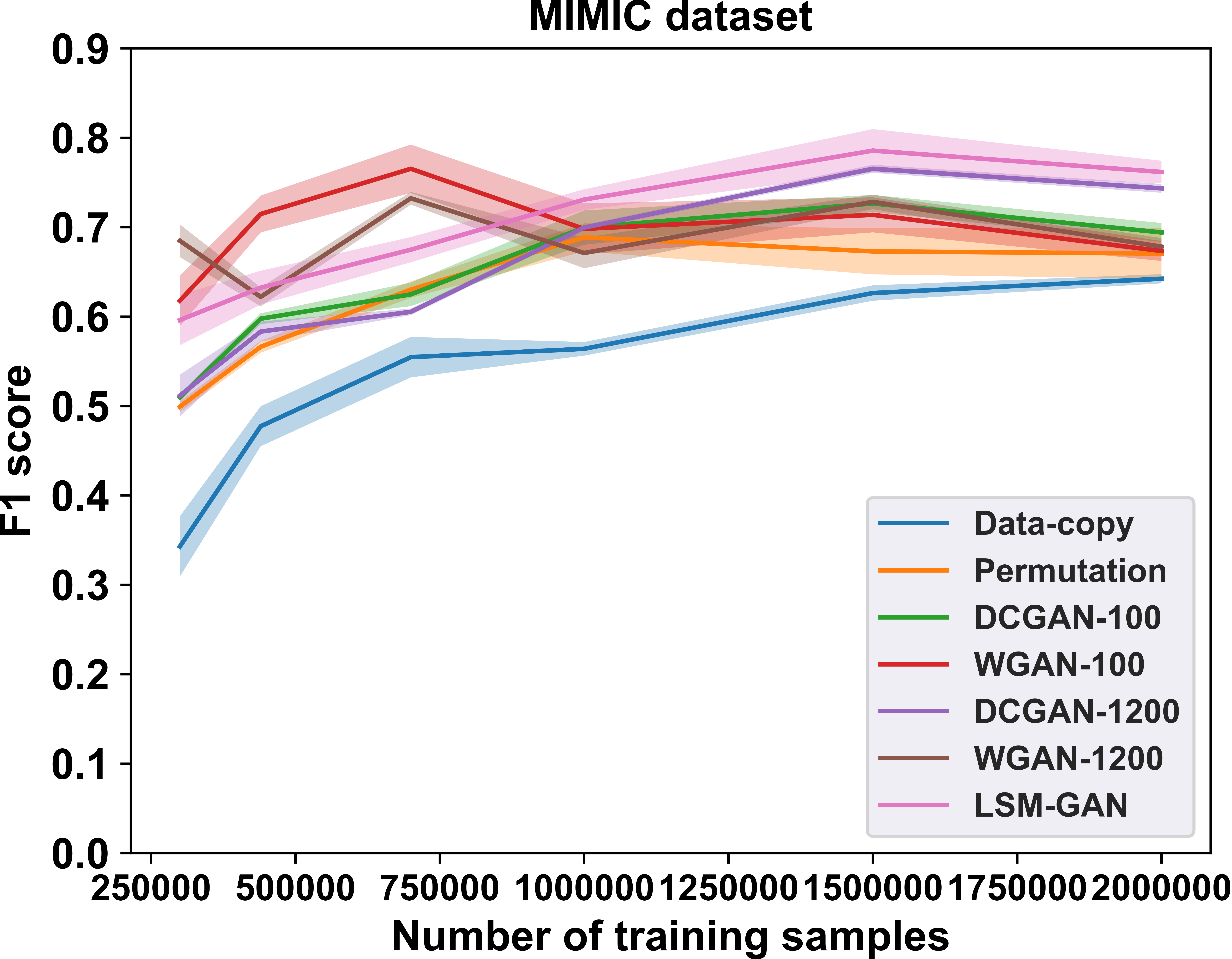}
\end{minipage}}
\caption{External validation of models from Experiment 3}
\label{fig8}
\end{figure}

A public dataset (DeepBeat dataset) released in 2020 \cite{19} with both signal quality and AF annotations was adopted to test the generalizability of the proposed approach. A data harmonization process was designed in the study, given the following three differences between our data and the DeepBeat data. First, the DeepBeat data is collected from wrist-type watches, while our data is collected from fingertips in the ICU setting. Second, the signal length of one segment is 25 seconds with a sampling rate of 32 Hz for DeepBeat, while ours is 30 seconds with a sampling rate of 240 Hz. Third, the preprocessing details were not reported clearly in the original paper and may differ from ours. Based on the above discrepancies in data, we first extended the 25-second segment to 30-second by stitching the first five seconds to the end of each signal (which may cause phase discontinuity). We then upsampled the signal length in the DeepBeat test set to the same as ours and adopted the min-max normalization on the data. Models from experiment 3 for each data augmentation method were selected to test on the DeepBeat dataset. A summary of performance can be found in Figure \ref{fig8} (a). Among our models, the proposed LSM-GAN \textbf{(pink curve)} consistently shows a leading performance in terms of F1 score, which is improved by 15\% compared to data-copying. 
The second external dataset was shared with us by authors in the work \cite{30}. They selected around 60 hours of PPG and synchronized ECG from 60 patients (containing both AF and non-AF patients) in the MIMIC-III waveform database. All the PPG signals were annotated by cardiologists from Guilin Medical University. However, PPG signals were segmented into 10-second strips in the study. To accommodate the dataset to our model, we repeat each 10-second signal two times, producing 30-second signals. Then same models from experiment 3 are tested on this dataset, results are reported in Figure \ref{fig8} (b). Although the proposed LSM-GAN does not lead at the early stage, it presents a consistently increasing trend with added samples and eventually arrives at the best F1 score across all data augmentation techniques. 

\subsection{Visualization}
\begin{figure}[ht]
\centering
\subfigure{
\begin{minipage}[t]{0.98\linewidth}
\centering
\includegraphics[width=1\columnwidth]{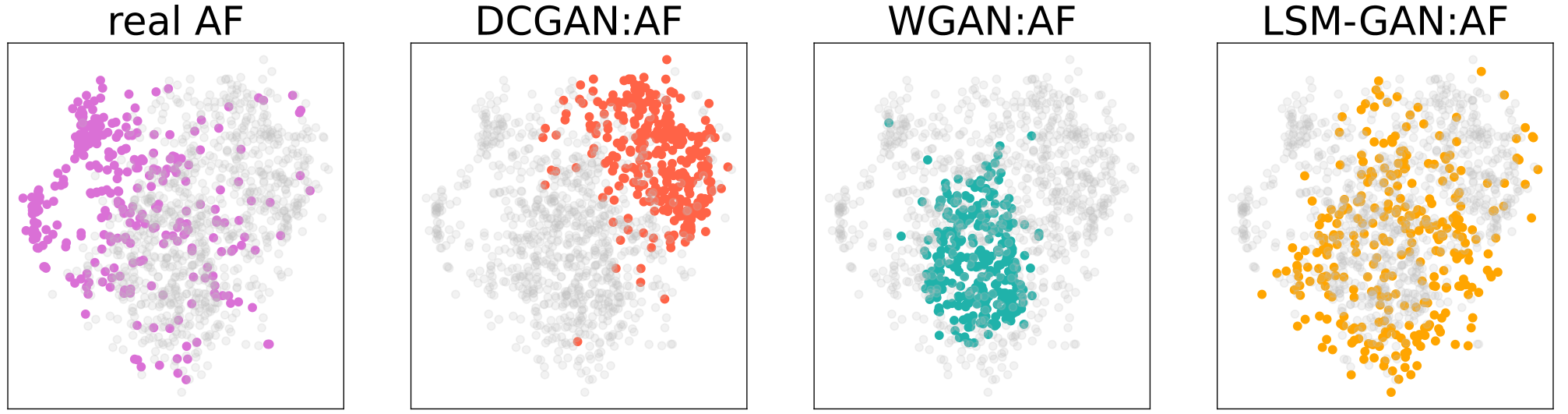}
\end{minipage}}\\
\subfigure{
\begin{minipage}[t]{0.98\linewidth}
\centering
\includegraphics[width=1\columnwidth]{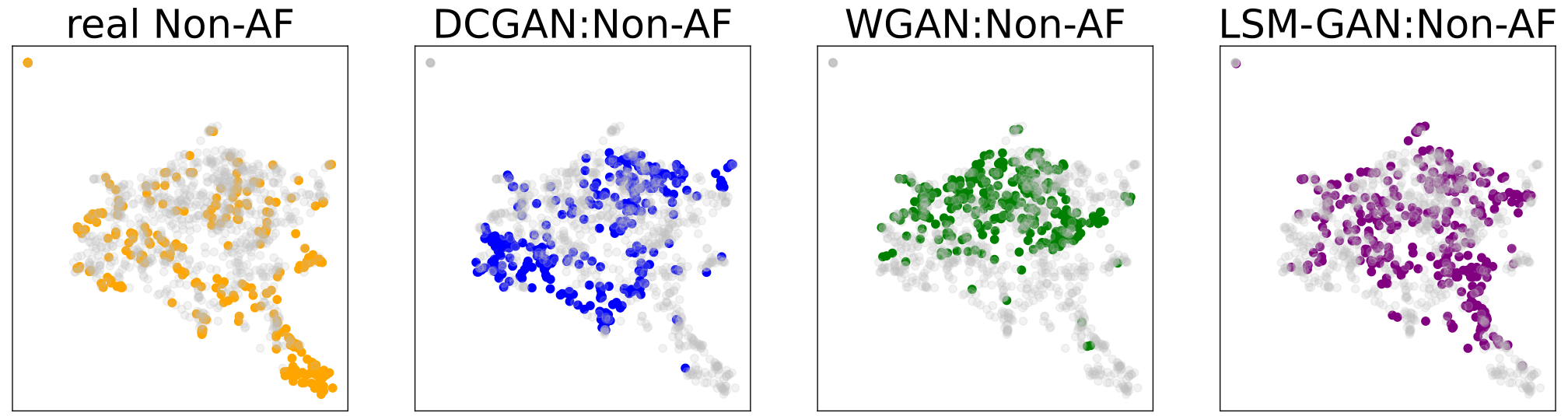}
\end{minipage}}
\caption{Visualization of distribution for real and synthetic AF signal.}
\label{fig9}
\end{figure}

 An additional experiment was conducted to visualize the distribution of the synthetic data generated by different GANs for both AF and Non-AF cases. We first randomly select 300 real signals and 300 synthetic signals from each GAN, then use Pairwise Controlled Manifold Approximation Projection (PaCMAP) \cite{31} to visualize the real signals together with synthetic signals in a 2D fashion. Figure \ref{fig9} presents the distribution of real and synthetic signals through PaCMAP calculated from signal amplitudes for both AF and Non-AF. Four sub-plots in each row contain the same dots, and each sub-plot only highlights dots from the related category.
Compared to the distribution of real AF signals, DCGAN and W-DCGAN only capture part of the distribution of real AF signals, and they do not share much overlap. On the other hand, the proposed LSM-GAN generates signals with similar distribution as real AF signals. Similar pattern can be observed in the Non-AF situation, DCGAN and WGAN only learned partial distribution of real Non-AF signals, while signals generated from LSM-GAN are more dispersed and distribute in a way similar to real ones.
\section{DISCUSSION}
\label{sec6}
 The proposed LSM-GAN integrates spectral information from the PPG waveform into the loss function to train generator in addition to the commonly used cross-entropy loss. Also, LSM-GAN uses a new architecture of generator which avoids up-sampling and aims to mimic a more well-understood filtering process to transform a white-noise into a narrow band signal like PPG. The analysis of the results is discussed as follows:

\subsection{Introducing spectral information in generated samples improves the AF detection accuracy}

Among all the data augmentation models reported in table \ref{tab2}, the proposed LSM-GAN achieves the best performance in accuracy and sensitivity with less than 1\% reduction on specificity and PPV. One plausible reason is that the randomness of phases of synthetic signals is allowed when matching in the spectral domain. Such randomness will enrich the training data and hence is beneficial for training AF models that analyze PPG in time domain including the ResNet architecture tested in this study This hypothesis can be partly supported by the PaCMAP visualization of generated signals from different data augmentation techniques in Figure \ref{fig9}, which reveals the PPG signals generated with LSM-GAN present a more similar distribution to the real signals than the other two GANs. 

\subsection{Increasing the dimension of random noise for GAN helps improve the AF detection accuracy}
We proposed a new generator architecture, which outputs the same length of the signal as the input (1200 in this study). To evaluate the effect of the new architecture, we added two more models: DCGAN-1200 and W-DCGAN-1200, which only changed the generator compared to the original DCGAN and W-DCGAN. Results from Experiment 1 (see Table \ref{tab2}) show that the new architecture improves 2\% in sensitivity, as evidenced by comparing DCGAN-100 to DCGAN-1200, and W-DCGAN-100 to W-DCGAN-1200. The key differentiator between the conventional architecture and the proposed one is whether an extra upsampling step is needed to ensure the length of a synthetic signal to be equal to a desired value. Learning GAN is essentially learning a series of transformations that convert random input noise into realistic synthetic data. In this study, when the generator does not alter the length of the input noise, the learned transformations can be better explained by a more well-understood filtering process. However, interpreting deep networks is still an ongoing effort and it remains interesting to uncover characteristics of the filters that are learned by LSM-GAN.

\subsection{Signal quality impacts effect of data augmentation}

As reported in Figure \ref{fig6}, increasing amount of artifacts in PPG signals has a negative influence on the AF detection performance. The AF model we used in this study - Resnet-50, is a generic model and does not have any specific modification to handle poor quality signals. Because our training dataset also contains imperfect PPG signal strips that are from both AF and non-AF classes, it is likely that a trained ResNet-50, based on such a training data set, will be confused when artifacts in the signal are learned as a pattern to be randomly associated with either AF or non-AF. However, the performance reduction can be alleviated through data augmentation methods. One plausible reason is because we only augmented good quality AF signals, in which the AF pattern is clear. Through data augmentation, real and clear AF patterns can be enhanced in the training data which reduce the negative influence of artifacts.
\subsection{LSM-GAN maintains superiority on external datasets}
In Figure \ref{fig8}, after evaluating our classifiers on two external datasets, we observe that our proposed LSM-GAN helps improve performance compared to using original data to train the model, and we can also observe the superiority of LSM-GAN compared to other data augmentation methods. However, there is a performance reduction on these datasets compared to results reported in the original publications. Exact reasons for this difference cannot be established without having access to the implemented algorithm as reported in the original studies but they are not the focuses of this study. However, because our study shows that performance of a generic deep neural network architecture as used in this study is particularly sensitive to the quality of a PPG signal, we speculate that DeepBeat database may contain large portion of poor quality PPG. Furthermore, DeepBeat algorithm explicitly incorporates PPG signal quality into AF detection and is expected to perform better on imperfect PPG signals. The original algorithm that was developed and tested on the MIMIC dataset did not consider PPG signal quality but it was developed to process 10-second PPG strips while our network was designed for processing 30-second PPG strips, which may be the likely reason for differences in performances between the two studies.  

\subsection{Limitations and future works}

Both AF and Non-AF PPG samples generated from LSM-GAN have a more similar distribution to that of real signals than samples generated from other data augmentation methods. However, what an adequate amount of data to train an effective GAN should be is not investigated. We only tested the effectiveness of LSM-GAN on the AF detection task, it remains interesting to test LSM-GAN on using PPG to detect other cardiac arrhythmia categories. Also, the present study focuses on the comparison of different DA techniques, so the same deep neural network architecture, i.e., ResNet-50, is adopted to achieve a fair comparison. This goal prevents us from exploring various deep learning models and customizations that may further improve the classification performance. Moreover, we can observe that from experiment 3 and external validation, although LSM-GAN boosts the performance, it still reaches a plateau of performance after one million training samples. This phenomenon indicates the limitation of data augmentation and that more real-world data is necessary to further improve the performance.

\section{CONCLUSIONS}
\label{sec7}

In the field of AI health, it remains difficult to obtain both large and well-annotated datasets. Data shortage and class-imbalance issues are standing challenges to properly train high-performing machine learning algorithms. In this study, we showed that properly designed GAN can potentially be used to augment and re-balance training data and improve classifiers solely trained on the original dataset that is imbalanced and contains fewer samples.

\bibliographystyle{ACM-Reference-Format}
\bibliography{sample-base}

\end{document}